\documentstyle[epsfig,mathptm,subfigure,amsmath]{mn}

\def\ltsima{$\; \buildrel < \over \sim \;$}
\def\gtsima{$\; \buildrel > \over \sim \;$}
\def\simlt{\lower.5ex\hbox{\ltsima}}
\def\simgt{\lower.5ex\hbox{\gtsima}}

\title[$H_0$ from an orientation-unbiased sample]{$H_0$ from an
orientation-unbiased sample of SZ and X-ray clusters}

\author[M.E. Jones et al.]{Michael E. Jones,$^{1}$ Alastair C. Edge,$^{2}$
 Keith Grainge,$^{1}$ William F. Grainger,$^{1}$ \cr R\"udiger
 Kneissl,$^{1}$ G.G. Pooley,$^{1}$ Richard Saunders,$^{1}$ Shigeru
 J. Miyoshi,$^{3}$ \cr Taisuke Tsuruta,$^{3}$ Koujun Yamashita,$^{4}$
 Yuzuru Tawara,$^{4}$ Akihiro Furuzawa,$^{4}$ \cr Akihiro Harada,$^{4}$
 Isamu Hatsukade$^{5}$\\ $^{1}$Astrophysics Group, Cavendish
 Laboratory, Madingley Road, Cambridge CB3 0HE\\ $^{2}$Department of
 Physics, South Road, Durham DH1 3LE \\ $^{3}$Department of Physics,
 Kyoto Sangyo University, Kamigamo-Motoyama, Kita-ku, Kyoto 603-8555,
 Japan \\ $^{4}$Department of Astrophysics, Faculty of Science, Nagoya
 University, Chikusa-ku, Nagoya 464-0814, Japan \\ $^{5}$Faculty of
 Engineering, Miyazaki University, 1-1 Gakuen-kibanadai-nishi,
 Miyazaki 889-2192, Japan}

\begin{document}

\maketitle

\begin{abstract}

We have observed the Sunyaev--Zel'dovich effect in a sample of five
moderate-redshift clusters with the Ryle Telescope, and used them in
conjunction with X-ray imaging and spectral data from {\sl ROSAT} and
{\sl ASCA} to measure the Hubble constant. This sample was chosen with
a strict X-ray flux limit using both the BCS and NORAS cluster
catalogues to be well above the surface-brightness limit of the {\sl
ROSAT} All-Sky Survey, and hence to be unbiased with respect to the
orientation of the cluster. This controls the major potential
systematic effect in the SZ/X-ray method of measureing $H_0$. Taking
the weighted geometric mean of the results and including the main
sources of random error, namely the noise in the SZ measurement, the
uncertainty in the X-ray temperatures and the unknown ellipticity of
the clusters, we find $H_0 = 59^{+8}_{-7} \rm \, km \, s^{-1} \,
Mpc^{-1}$ assuming a standard CDM model with $\Omega_{\rm M} = 1.0,
\Omega_{\Lambda} = 0.0$, or $H_0 = 65^{+8}_{-7} \rm \, km \, s^{-1} \,
Mpc^{-1}$ if $\Omega_{\rm M} = 0.3, \Omega_{\Lambda} = 0.7$.

\end{abstract} 


\section{Introduction}

It has long been recognised that the Sunyaev--Zel'dovich (SZ) effect
\cite{s&z} in clusters of galaxies, along with X-ray measurements, can
provide a method of measuring distances on cosmological scales
\cite{G78,SW78,cavaliere,B79}. The method takes advantage of the
differing dependences on the cluster electron density $n_{\rm e}$ and
temperature $T_{\rm e}$ of the X-ray bremsstrahlung ($\propto n^2_{\rm
e} T^{1/2}_{\rm e}$) and SZ effect ($\propto n_{\rm e} T_{\rm e}$) to
measure the physical size of the cluster, and hence its
angular-diameter distance. The dependence of the derived value of
$H_0$ on the important quantities is
\[
H_0 \propto \left(\frac{T_{\rm e}}{\Delta T_{\rm SZ}}\right)^2\, \theta \, S_{\rm X} \frac{\ell_{\bot}}{\ell_{\|}},
\]
where $\Delta T_{\rm SZ}$ is the SZ brightness decrement, $S_{\rm X}$
is the X-ray surface brightness, $\theta$ is the apparent angular size
of the cluster, and $\frac{\ell_{\bot}}{\ell_{\|}}$ is the ratio of
the linear sizes of the cluster perpendicular and parallel to the line
of sight.

The systematic effects in this method of measuring $H_0$ are quite
different from those in other methods, either in the traditional
distance ladder or in other direct physical methods such as
gravitational lensing, and are mostly due to the difficulty of
modelling the cluster gas that causes both the SZ effect and the X-ray
emission. Recent work has shown that the small-scale features in the
gas distribution such as cooling flows or clumping contribute smaller
biases than might have been thought, since the effects on the observed
quantities entering $H_0$ tend to cancel each other even for a single
cluster (eg, Grainger \shortcite{Gr01th}, Maggi et al.). However, the
large-scale distribution of the gas poses a problem, since the
line-of-sight depth through the cluster cannot be measured directly,
and has to be assumed to be the same as the size of the cluster in the
plane of the sky.  This asumption will be violated if clusters are
chosen preferentially because of high surface brightness, since this
will favour clusters that are extended along the line of sight. This
will bias low the derived value of $H_0$. We therefore need a cluster
sample that is selected on total flux rather than surface brightness,
well above the surface brightness limit of the survey. Although
several individual $H_0$ estimates from the SZ/X-ray method have been
published (eg \cite{HB98,H,R00}), properly selected cluster samples
are only recently beginning to emerge \cite{mason2001}. In this paper
we derive a sample of moderate-redshift X-ray-selected clusters that
is carefully controlled to be free of any orientation bias, and use it
to calculate $H_0$ using the SZ data from the Ryle Telescope (RT).

Unless stated otherwise, we assume a standard cold dark matter
cosmology (SCDM, $\Omega_{\rm M} = 1.0, \, \Omega_{\Lambda} =
0.0$). We also give our results assuming a $\Lambda$CDM cosmology
($\Omega_{\rm M} = 0.3, \, \Omega_{\Lambda} = 0.7$).

\section{X-ray cluster selection}

The most complete X-ray cluster samples suitable for our purpose are
selected from the {\sl ROSAT} All-Sky Survey (RASS). We use two such
samples, the Bright Cluster Sample (BCS) \cite{BCS}, and Northern
Rosat All-Sky Survey (NORAS) \cite{NORAS}. We began selection of our
SZ sample from BCS while it was being compiled; we use NORAS as a
consistency check. Both surveys are based on cluster candidates found
in the first processing (SASS) of the RASS data, with further tests to
check they are indeed clusters. A problem, particularly for clusters
at high redshift, is the risk of incompleteness for compact clusters
that may not appear extended enough to be recognised as such. BCS
attempts to overcome this problem by adding known Abell and Zwicky
clusters to the sample. However, comparing the BCS and NORAS samples
after applying our RT selection criteria, it turns out that our sample
is not affected by differences between the BCS and NORAS secondary
selections.

\begin{figure}
\centerline{\epsfig{file=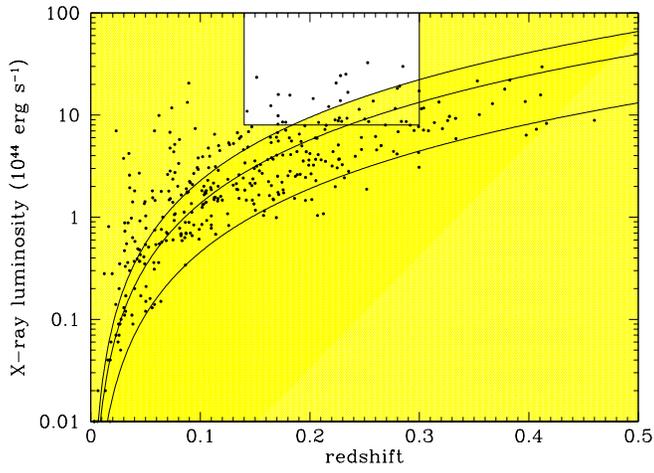,angle=-90,width=9cm}}
\caption{The full NORAS X-ray cluster sample and the region (white) of
our SZ subsample. The lines correspond to flux limits of 1,~3~and $5
\times 10^{-12}$ erg s$^{-1}$ cm$^{-2}$, which are the approximate
lower detection limit, the 90\% completeness limit, and the limit of
the SZ sample respectively. }
\label{fig:noras_sample}
\end{figure}

The flux limit of NORAS is $3 \times 10^{-12}$ erg s$^{-1}$ cm$^{-2}$
at 90\% completeness; that of BCS is $4.4 \times 10^{-12}$ erg
s$^{-1}$ cm$^{-2}$ to the same completeness. But we limit our sample
to even higher fluxes, of greater than $5 \times 10^{-12}$ erg
s$^{-1}$ cm$^{-2}$ (see Figure~\ref{fig:noras_sample}). Both this and
a restriction to redshifts $z<0.3$ help reduce the risk of missing
compact clusters. In accordance with the requirements for the Ryle
Telescope we restrict our sample to the northern sky at declinations
$\delta >$ 20$^\circ$, to a redshift range of $z > 0.14$, and
luminosities of $L_x > 8 \times 10^{44}$ erg s$^{-1}$. Table
\ref{table:sample} lists the clusters from BCS and from NORAS that
meet these criteria, and also indicates those which have radio sources that
would make SZ measurement difficult or impossible. Where the surveys
disagree on whether a cluster should be included on grounds of flux,
we cautiously select only those clusters that meet our criteria in
both surveys: this excludes three clusters above our flux limits in
BCS and not in NORAS, and one in NORAS but not in BCS. This is
consistent with the errors in the flux estimates in each survey. Two
BCS-selected clusters are not in NORAS, consistent with the inclusion
of optically selected clusters in BCS that are not in NORAS due to
lack of extent. Of the eleven clusters satisfying our X-ray criteria,
six have radio sources in the cluster or within the field of view of
the RT that are too bright at 15~GHz to allow SZ observation with the
RT.  We have observed the remaining five, A697, A773, A1413, A1914 and
A2218 with the RT and obtained detections for all of them; the results
are discussed below. Since the presence of radio sources cannot be
correlated with cluster orientation, this subsample is effectively
randomly chosen, and therefore retains the orientation-independence of
the whole sample. Indeed, we have checked that the our subsample is not
correlated to X-ray properties such as flux, extent likelihood and
measured extent. It therefore appears unlikely that any significant
bias, larger than the random error, is present in our averaged $H_0$
result. Clearly a larger sample would improve matters: the small size
of our sample highlights the difficulty of selecting rigorously at
moderately high redshift, even using the best large-area X-ray surveys
available.

\section{SZ observations and $H_0$ calculation}

Details of our observational and modelling methods are given in
Grainge et al.~\shortcite{G01a}, Grainge et al. \shortcite{G01b} and
Grainger et al. \shortcite{Gr01}. We briefly summarise them here.

In each case we fit the {\sl ROSAT} PSPC or HRI image using an
ellipsoidal $\beta$-model, with $\beta$, the core radii in the plane
of the sky, $\theta_1$ and $\theta_2$, the position angle of the major
axis, the position on the sky, and the central electron density $n_0$
all as free parameters. The core radius in the line of sight is set to
the geometric mean of the other two. A downhill simplex method is used
to maximise the likelihood of the data given the model, using Poisson
statistics. For the PSPC images we use the hard band (0.5--2 keV) in
order to minimise the effect of Galactic absorption.

We remove contaminating radio sources by a maximum likelihood fit to
the visibility data, simultaneously fitting the SZ decrement (using
the X-ray-determined model, with a variable amplitude) and the
sources. We include trial sources at all positions suggested either
from the long baseline RT data or from other observations such as NVSS
\cite{NVSS}, and allow their positions and fluxes to vary in the
fit. This allows us to remove the effects of sources whose existence
is known from lower-frequency observations but whose brightness would
not warrant a significant detection from the RT data alone.

Finally we take the source-subtracted data and compare them with mock
SZ data derived from the X-ray model, taking into account the full
response of the RT. We vary the normalisation of the mock data
(equivalent to varying $H_0$) and find the likelihood of the data, and
then multiply by a prior that is uniform in log space (since $H_0$ is
a scale parameter) to find the peak and extent of the posterior
probability distribution.  We then apply corrections for the more
exact relativistic form of the Compton $y$ parameter
\cite{challinor98}, and small time-dependant variations from the
nominal RT flux calibration scale (determined from VLA monitoring of
our primary flux calibrators, 3C48 and 3C286), as well as adding in
(in quadrature) estimates of the other main sources of error.

We now give details of the procedure for each of the five
clusters. Two have been published more fully elsewhere (A773, Saunders
et al. \shortcite{S01} and A1413, Grainge et al. \shortcite{G01b}) and
are briefly summarised here; one (A2218) is a re-analysis of
previously published data \cite{jones-capri}; the other two (A697 and
A1914) are new.

\subsection{A697}

\begin{figure*}
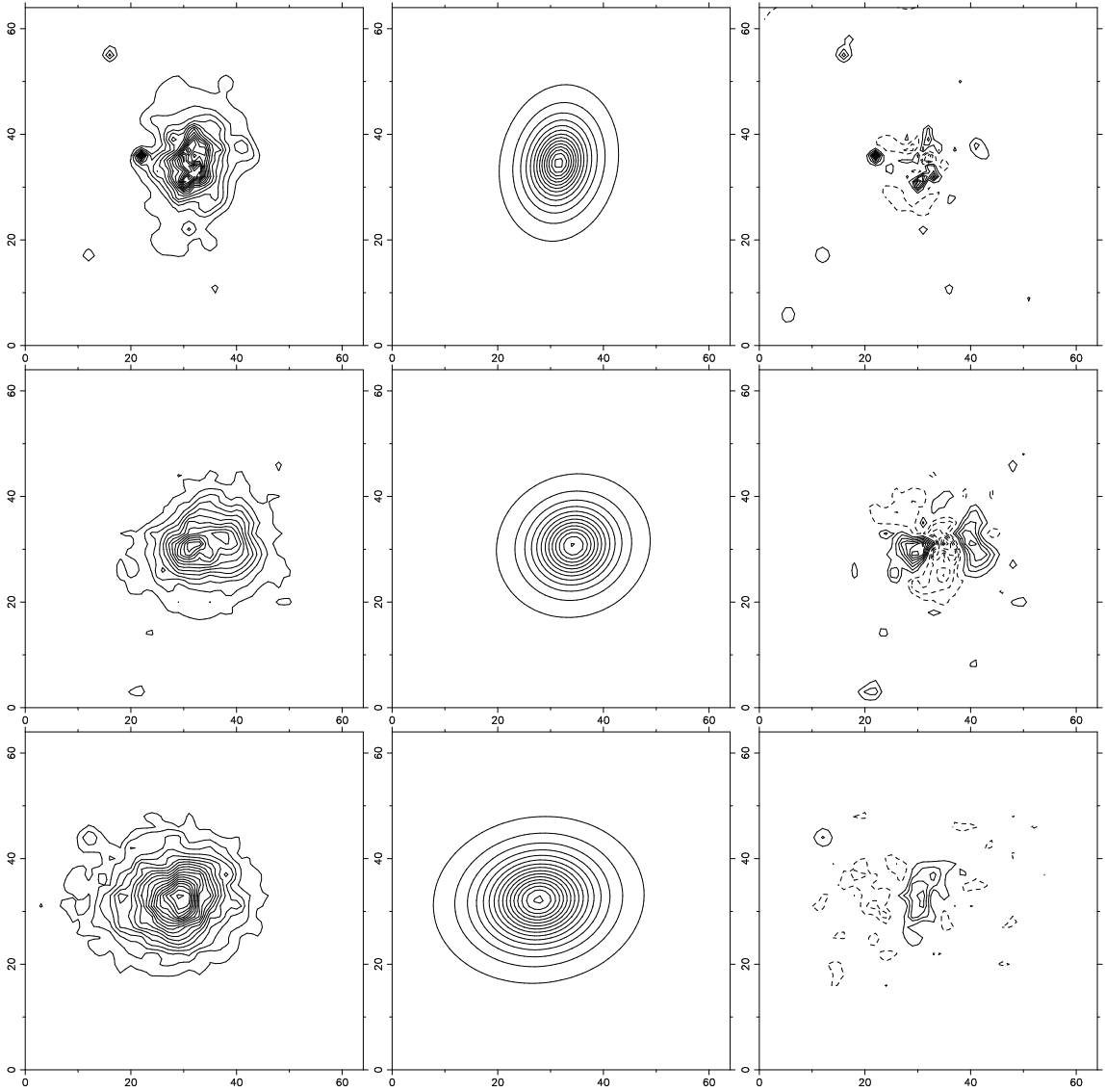

\hbox{
\epsfig{file=A697-rosat.eps,angle=-90,width=5cm}
\epsfig{file=A697-model.eps,angle=-90,width=5cm}
\epsfig{file=A697-residual.eps,angle=-90,width=5cm}
}
\hbox{
\epsfig{file=A1914-rosat2.eps,angle=-90,width=5cm}
\epsfig{file=A1914-model2.eps,angle=-90,width=5cm}
\epsfig{file=A1914-resi2.eps,angle=-90,width=5cm}
}
\hbox{
\epsfig{file=A2218-rosat.eps,angle=-90,width=5cm}
\epsfig{file=A2218-model.eps,angle=-90,width=5cm}
\epsfig{file=A2218-residual.eps,angle=-90,width=5cm}
}
\caption{X-ray image, best-fitting model, and residuals for (top)
A697, (centre) A1914 and (bottom) A2218. The contour intervals are
respectively 1 count per 4 arcsec pixel, 2 counts per 8 arcsec pixel
and 2 counts per 8 arcsec pixel.}
\label{fig:xray-model}
\end{figure*}

A697 was observed with the {\sl ROSAT} HRI on 1995 November 6 with a
live time of 28.1 ks. We also observed it using ASCA during 1996 April
in order to determine its temperature. Analysing both the GIS and SIS
data using standard XSPEC tools, we find a temperature of
$8.0^{+0.6}_{-0.4}$ keV (68\% confidence), with a metallicity of $0.24
\pm 0.07$ solar. Fitting to the {\sl ROSAT} HRI data, we find the
parameters given in Table \ref{table:xray}. The HRI image, model and
residuals are shown in Figure \ref{fig:xray-model}.

\begin{figure*}
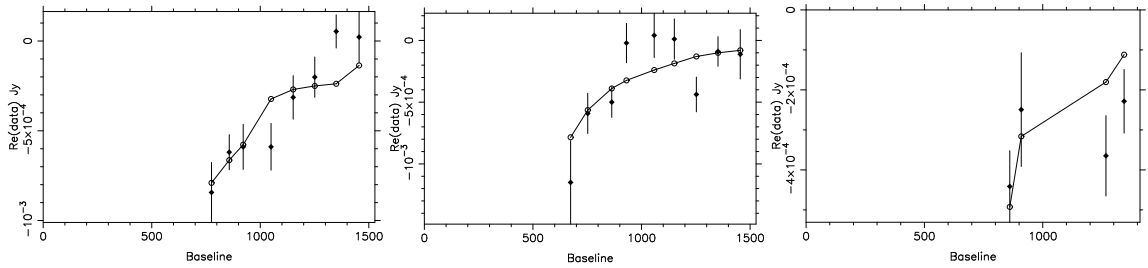

\hbox{
\epsfig{file=a697-fit.eps, angle = -90, width=5cm}
\epsfig{file=a1914-fit2.eps, angle = -90,width=5cm}
\epsfig{file=a2218-fit.eps, angle = -90,width=5cm}
}
\caption{Source-subtracted visibilities (real part only) for (left)
A697, (centre) A1914 and (right) A2218 (filled points with error
bars), binned azimuthally, compared with the best-fitting model in
each case (open points joined by lines). The lack of smoothness in the
models is due to the inclusion of points from different baselines with
different model values (due to ellipticity) in the same bin.}
\label{fig:visplots}
\end{figure*}

\begin{figure*}
\hbox{
 \epsfig{file=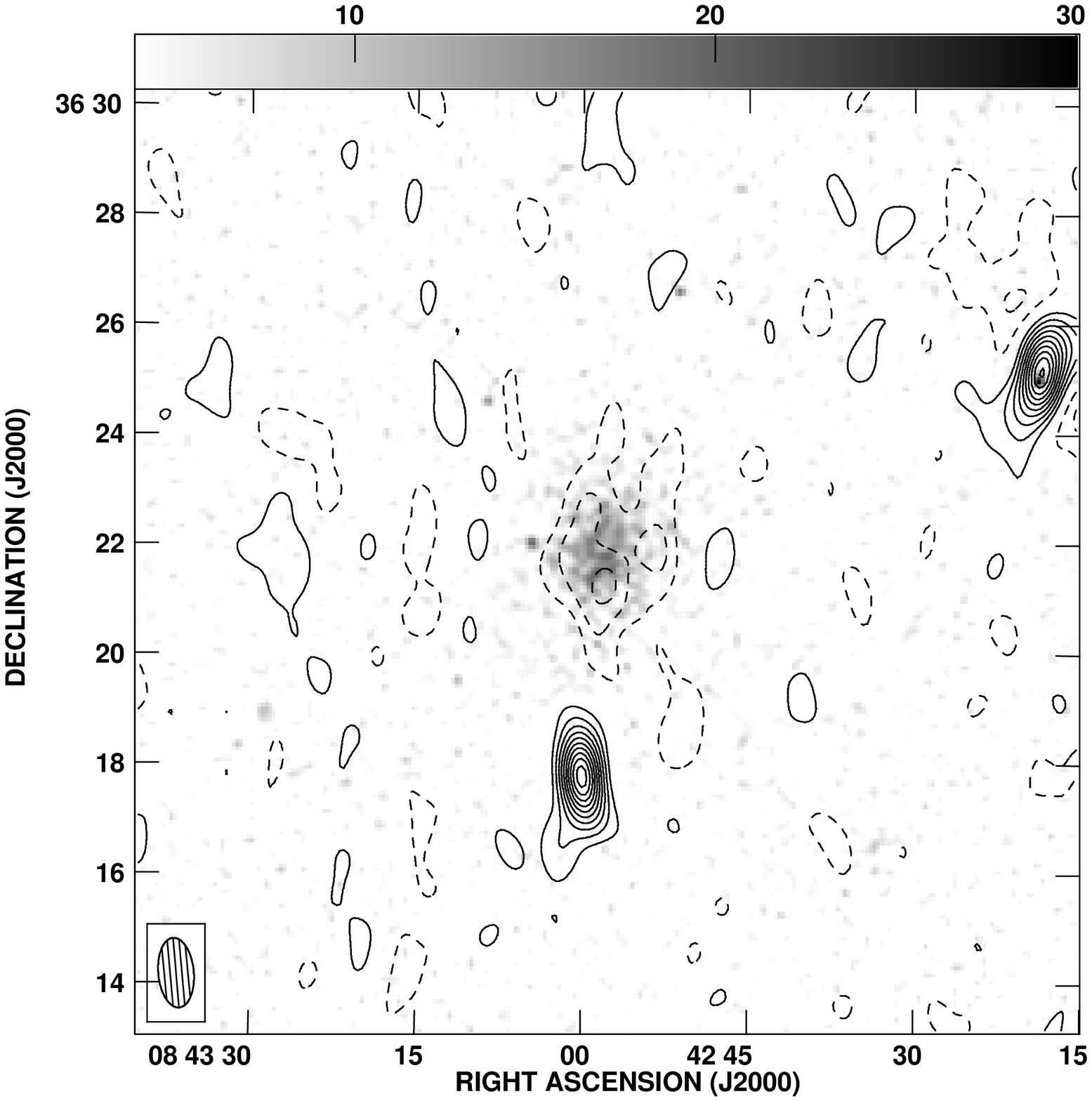,width=8cm}
 \epsfig{file=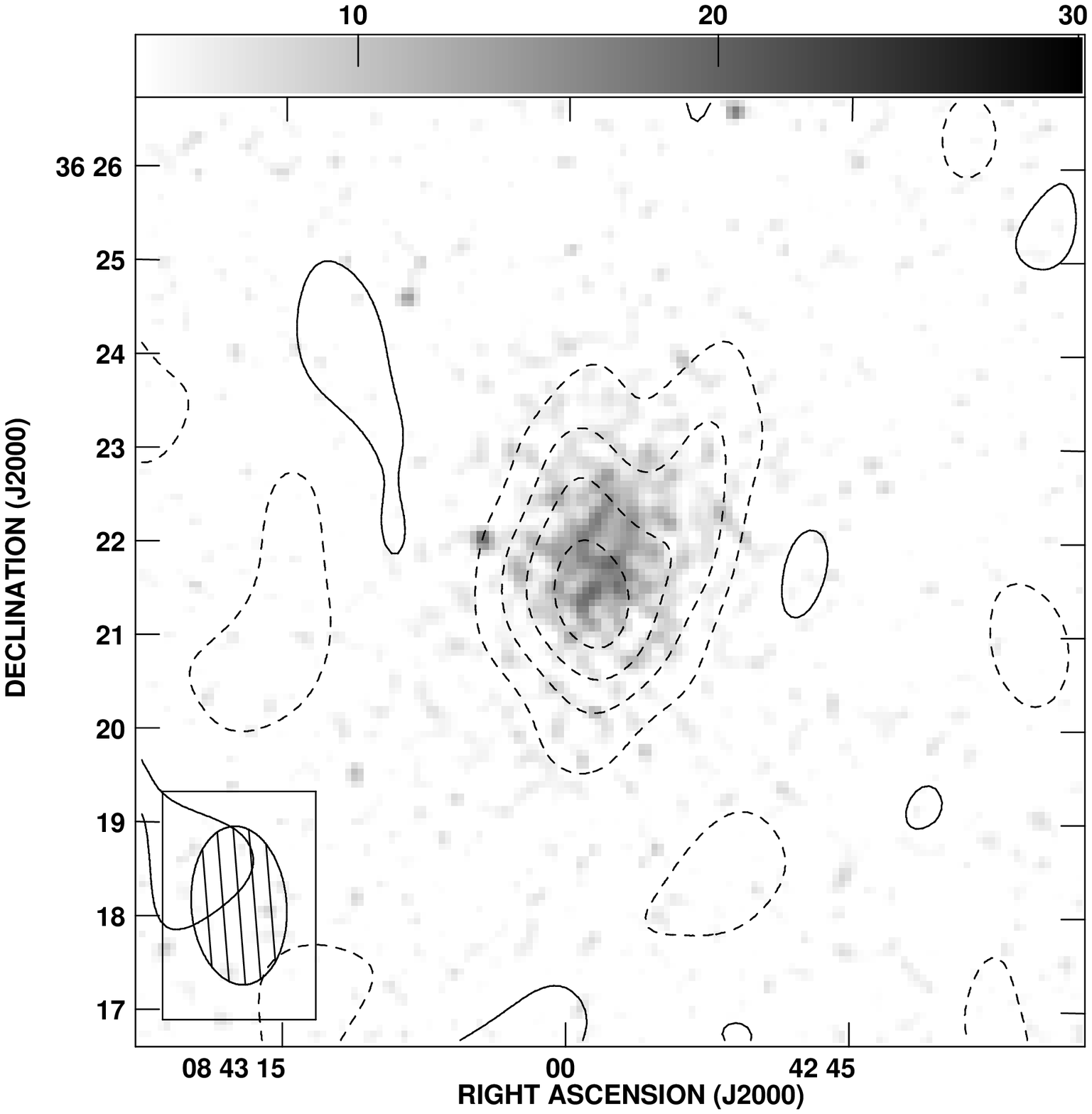,width=8cm}
}
\caption{(Left)CLEANed map of all the radio data on A697 (contours)
superposed on the {\sl ROSAT} HRI image (greyscale). The contour
interval is $60 \, \mu \rm Jy \, beam^{-1}$; dashed contours are
negative. The CLEAN restoring beam of $101^{\prime \prime} \times
61^{\prime \prime}$ is shown (inset). (Right)CLEANed map of the
source-subtracted radio data on A697, with a gaussian taper of width
$2000 \lambda$ applied (contours) superposed on the {\sl ROSAT} HRI
image (greyscale). The contour interval is $60 \, \mu \rm Jy \,
beam^{-1}$; dashed contours are negative. The CLEAN restoring beam of
$76^{\prime \prime} \times 39^{\prime \prime}$ is shown (inset).}
\label{fig:A697}
\end{figure*}

We observed A697 with the RT on 42 occasions between 1994 August 6 and
1996 April 27.  The map of all the data is shown in Figure
\ref{fig:A697}. The bright source to the NW is extended, but can be
successfully modelled as three point sources. Fitting simultaneously
to the sources and SZ decrement gives a good fit with the source
parameters listed in Table \ref{table:sources}. Fitting the
source-subtracted data to the model derived from the X-ray data
(Figure \ref{fig:visplots}), and applying the relativistic and
flux-scale corrections, we find $H_0 = 52^{+8}_{-6} \rm \, km \, s^{-1}
\, Mpc^{-1}$ (errors from SZ fitting only). Including the other
errors, dominated by the X-ray temperature ($^{+14}_{-10}\, \%$) and
ellipticity ($\pm 14\%$), we find $H_0 = 52^{+13}_{-11} \rm \, km \,
s^{-1} \, Mpc^{-1}$. For a $\Lambda$CDM cosmology with $\Omega_{\rm M}
= 0.3$ and $\Omega_{\Lambda} = 0.7$, this becomes $H_0 =
59^{+15}_{-12} \rm \, km \, s^{-1} \, Mpc^{-1}$.

\subsection{A773}

Details of the determination of $H_0$ from A773 are given in Saunders
et al. \shortcite{S01}. We used a {\sl ROSAT} HRI image and a temperature
measurement from ASCA in conjunction with our RT data to estimate a
value of $H_0 = 77^{+19}_{-17} \rm \, km \, s^{-1} \, Mpc^{-1}$ for SCDM,
$H_0 = 85^{+20}_{-19} \rm \, km \, s^{-1} \, Mpc^{-1}$ for $\Lambda$CDM.

\subsection{A1413}

Details for A1413 are given by Grainge et al. \shortcite{G01b}, who
also discuss the contributions to the $H_0$ error budget in
general. Using a {\sl ROSAT} PSPC image and ASCA temperature, we find
$H_0 = 57^{+23}_{-16} \rm \, km \, s^{-1} \, Mpc^{-1}$ for SCDM, $H_0 =
61^{+25}_{-17} \rm \, km \, s^{-1} \, Mpc^{-1}$ for $\Lambda$CDM.

\subsection{A1914}

\begin{figure*}
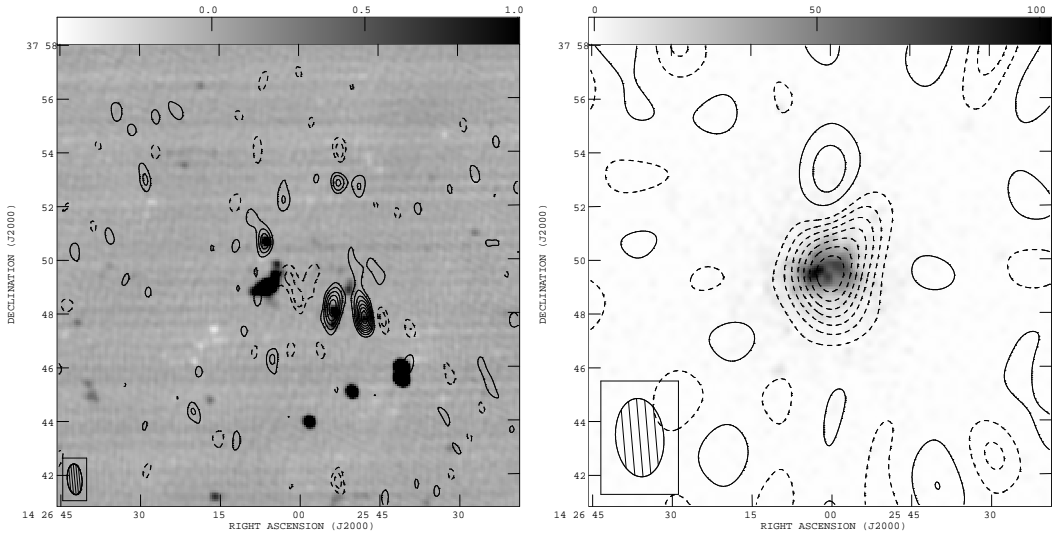

\hbox{ \epsfig{file=A1914-hires.ps, width=7cm}
\epsfig{file=A1914-lores.ps, width=7cm} }
\caption{(left) Full-resolution radio image of A1914 (contours)
overlaid on a VLA 1.4~GHz image (greyscale). The contours are at $-3,
-2, 2, 3, 4,5 , 6,7 , 8 \times 35\, \mu \rm Jy \, beam^{-1}$,; the
beamsize of $69 \times 33$ arcsec, PA $5^{\circ}$ is shown
(inset). The steep-spectrum halo source can be seen in the VLA image
just to the east of the cluster centre -- no emission is apparent in
the RT image however, where the SZ decrement can also be seen. (right)
Source-subtracted CLEANed low-resolution radio image of A1914 showing
the SZ decrement, overlaid on the {\sl ROSAT} PSPC image
(greyscale). The contour interval is 80 $\mu\, \rm Jy \, beam^{-1}$,
dashed contours are negative. The beam (inset) is $175'' \times
108''$, PA $4^{\circ}$.}
\label{fig:A1914}
\end{figure*}

A1914 was observed with the {\sl ROSAT} PSPC for 8.6 ks in 1992
July. There are several pieces of evidence to suggest that A1914 is
undergoing a merger. The {\sl ROSAT} image (see Figure
\ref{fig:A1914}) shows significant substructure in the core of the
cluster, but no evidence of a cooling flow. The temperature, measured
from an ASCA observation, is rather high at $8.56 \pm 0.39$ keV. The
POSS optical image shows two distinct groups of galaxies with no
single dominant galaxy, and finally there is a diffuse steep-spectrum
radio source to the NE of the cluster centre that is detected in the
WENSS (300 MHz) \cite{WENSS} and NVSS (1.4 GHz) surveys. This raises
the question of whether A1914 should be included in our sample, since
it is possible that the cluster gas might be far from hydrostatic
equilibrium and difficult to model. We choose to include it in our
sample since simulations \cite{Gr01th} show that even including
clusters with large internal kinetic energies in their gas does not
bias the sample average $H_0$ However, such clusters can have a large
scatter about the mean value -- this does appear to be the case with
A1914 and also with A2218.

We fitted a $\beta$ model to the {\sl ROSAT} PSPC image with the
parameters given in Table \ref{table:xray}. The residuals (see Figure
\ref{fig:xray-model}) were noticeably poorer than those from the fits
to other clusters such as A1413; a Monte Carlo analysis shows that the
hypothesis that the true structure is a $\beta$ model and that the
features seen result from Poisson noise can be rejected at the
4-$\sigma$ level. A1914 was observed with the RT on 19 occasions
between 1996 March 6 and April 16; the full resolution radio image is
shown in Figure \ref{fig:A1914}. There are clearly several radio
sources near the cluster centre, some of which are also detected in a
VLA image (Figure \ref{fig:A1914}). We used the VLA source
positions and positions from a long-baseline RT image to estimate the
fluxes of nine sources; the fluxes found and subtracted are shown in
Table \ref{table:sources}. Fitting the source-subtracted
short-baseline data to the X-ray-based model (Figure
\ref{fig:visplots}) results in a value of $H_0 = 119^{+42}_{-31} \rm
km \, s^{-1} \, Mpc^{-1}$ with errors just from the SZ fitting, and
$H_0 = 119^{+46}_{-38} \rm \, km \, s^{-1} \, Mpc^{-1}$ including the
other sources of error. For $\Lambda$CDM this becomes $H_0 =
129^{+51}_{-42} \rm \, km \, s^{-1} \, Mpc^{-1}$.

\subsection{A2218}

A2218 was the first cluster to be observed in the RT SZ programme
\cite{J93}, and has also been extensively observed in many wavebands,
with SZ detections in the radio (eg Birkinshaw \&
Hughes~\shortcite{B94}) as well as strong and weak lensing
observations (eg Kneib et al.~\shortcite{K95} and Squires et
al.~\shortcite{S96}) and X-ray studies (eg
Markevitch~\shortcite{M97}). We have previously reported a measurement
of $H_0$ from this cluster \cite{jones-capri}; we re-analyse our data
here because we have improved our source-subtraction and modelling
techniques, and because a new temperature measurement from ASCA is
available that supersedes the previous GINGA measurement.

\begin{figure*}
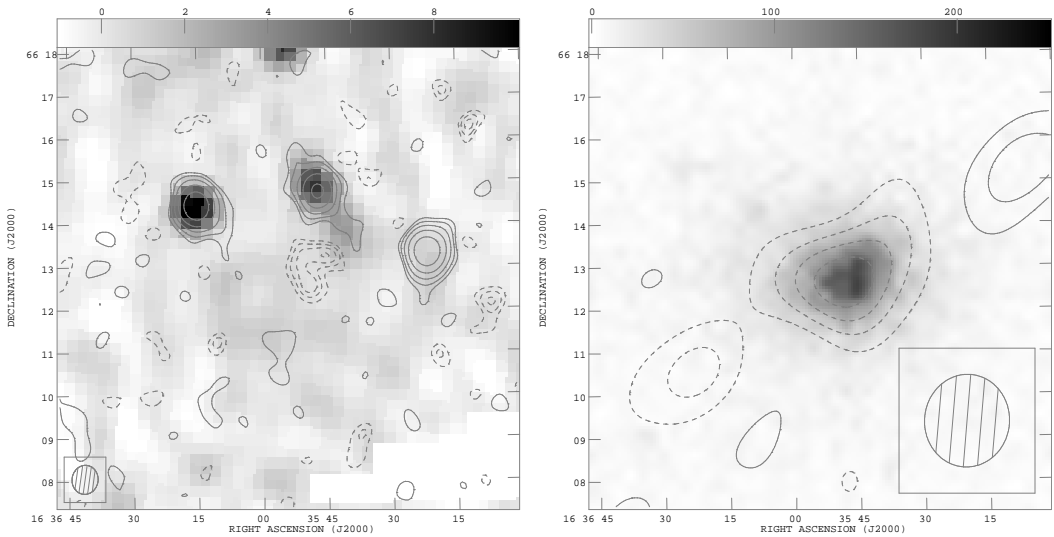

\hbox{
\epsfig{file=A2218-hires.ps, width=7cm}
\epsfig{file=A2218-lores.ps, width=7cm}
}
\caption{(left) Full-resolution radio image of A2218 (contours)
overlaid on the NVSS 1.4~GHz image (greyscale). The contours are at
$-4, -3, -2, 2, 4, 8, 16, 32 \times 35\, \mu \rm Jy \, beam^{-1}$; the
beamsize of $41 \times 36$ arcsec, PA $-9^{\circ}$ is shown
(inset). The SZ decrement can be clearly seen even in this
high-resolution image. (right) Source-subtracted CLEANed
low-resolution radio image of A2218 showing the SZ decrement, overlaid
on the {\sl ROSAT} PSPC image (greyscale). The contours interval is
$80 \,\mu\, \rm Jy \, beam^{-1}$, dashed contours are negative. The
beam (inset) is $130'' \times 120''$, PA $-5^{\circ}$.}
\label{fig:A2218}
\end{figure*}

Figure \ref{fig:A2218} shows the image made by combining 30
days' RT data taken in two array configurations between 1992 December
4 and 1993 March 23. This map has simply been CLEANed with no source
subtraction; nevertheless, even at this high resolution the SZ
decrement is clearly visible, and an approximate calculation of the
observed temperature decrement, using the CLEAN restoring beam size of
$40^{\prime \prime} \times 36^{\prime \prime}$, gives a maximum
decrement of $-760 \pm 150 \, \mu$K. This agrees well with the fitted
$\beta$-model central temperature of $-797\, \mu$K.

We used the {\sl ROSAT} PSPC observation made on 1991 May 25, which
has a live time of 42.47~ks. The X-ray model fitting parameters are
shown in Table \ref{table:xray}. As with A1914, the X-ray image
contains substructure that is significantly different from a smooth
ellipsoidal $\beta$ model (see Figure
\ref{fig:xray-model}). Preliminary source subtraction in the map plane
had shown that, of the three bright sources evident in Figure
\ref{fig:A2218}, the central one was point-like but the eastern
and western sources had some additional extended flux. We therefore
fitted a total of six trial sources, modelling the western source as
three points and the eastern source as two. This gave a good fit to
the data with the source parameters listed in Table
\ref{table:sources}. Subtracting these sources and fitting to the
X-ray model (using an ASCA-derived temperature of $7.20 \pm 0.36$ keV
\cite{a2218-temp}), we find a best fit value of $H_0 = 31_{-6}^{+8}
\rm \, km \, s^{-1} \, Mpc^{-1}$, where these errors only include the SZ
fitting. Including the other dominant sources of error ($\pm 10 \%$
from the temperature measurement and $\pm 14 \%$ from the likely
ellipticity), we find $H_0 = 31^{+10}_{-8}\rm \, km \, s^{-1} \,
Mpc^{-1}$, or $H_0 = 35_{-9}^{+11} \rm \, km \, s^{-1} \, Mpc^{-1}$ for
$\Lambda$CDM.

\subsection{Combined result}

The values of $H_0$ derived from each cluster are summarised in in
Table \ref{table:h0}. Since the main sources of error in each value
are in the form of a multiplicative factor, we calculate the sample
mean by taking a weighted average of $\log H_0$. The result is $H_0 =
59^{+8}_{-7} \rm \, km \, s^{-1} \, Mpc^{-1}$ for SCDM or $H_0 =
65^{+8}_{-7} \rm \, km \, s^{-1} \, Mpc^{-1}$ for
$\Lambda$CDM. Looking at the scatter in our individual values, we see
that A697, A773 and A1413 all agree with the mean within
1-$\sigma$. A1914 is 1.6 $\sigma$ above the mean and A2218 is 2.8
$\sigma$ below it. These are the two clusters for which there is
independent evidence, from the lack of a good fit of the X-ray model
and from the optical galaxy distribution, of significant dynamical
activity. Since our error estimate does not include a component to
take account of this effect (for lack of a means of quantifiying it),
it is not suprising that the scatter appears anomalously
large. Nevertheless we feel justified in keeping these two clusters in
the sample based on the evidence from simulations that dynamically
active clusters do not bias the value of $H_0$ -- indeed, removing
them does not significantly change the best estimate of $H_0$.

\section{Conclusions}\label{conclusions}

We have used {\sl ROSAT} and ASCA X-ray data and Ryle Telescope SZ
observations to measure the Hubble constant $H_0$ from a sample of
five clusters, selected using the BCS and NORAS cluster catalogues to
be free of orientation bias. The weighted geometric mean value of
$H_0$ from this sample is $H_0 = 59^{+8}_{-7} \rm \, km \, s^{-1} \,
Mpc^{-1}$ assuming a standard CDM model with $\Omega_{\rm M} = 1.0,
\Omega_{\Lambda} = 0.0$, or $H_0 = 65^{+8}_{-7} \rm \, km \, s^{-1} \,
Mpc^{-1}$ if $\Omega_{\rm M} = 0.3, \Omega_{\Lambda} = 0.7$.

\section{Acknowledgements}

We thank the staff of the Cavendish Astrophysics group who ensure the
operation of the Ryle Telescope, which is funded by PPARC. AE
acknowledges support from the Royal Society; RK acknowledges support
from an EU Marie Curie Fellowship; WFG acknowledges support from a
PPARC studentship. We have made use of the ROSAT Data Archive of the
Max-Planck-Institut f\"ur extraterrestrische Physik (MPE) at Garching,
Germany. We thank Hans B\"ohringer for helpful discussions.

\clearpage
\begin{table*}
\begin{tabular}{ccccccccccc}
Cluster&R.A. (J2000)& Dec &$z$&$f_{\rm X}^{\rm BCS}$&$L_{\rm X}^{\rm BCS}$&$f_{\rm X}^{\rm NORAS}$&$L_{\rm X}^{\rm NORAS}$& B?& N?& R? \\
\hline	
 A586 &   113.091 &   31.629 & 0.1710 &   9.1 & 11.12  &     7.94 &  9.84  &  *  &    * &   \\
 A665 &   127.739 &   65.854 & 0.1818 &  11.8 & 16.33  &    11.23 & 15.69  &  *  &    * &   \\
{\bf A697} &   130.741 &   36.365 & 0.2820 &   5.0  &16.30   &    5.77 & 19.15  &  *  &    * & * \\
{\bf A773} &   139.475 &   51.716  &0.2170 &   6.7 & 13.08  &     5.71 & 12.11  &  *  &    * & * \\
 Z2701&   148.198  &  51.891 & 0.2140 &   5.6 & 10.68  &      --  &   --   &  *  &      &   \\
 A963 &   154.255 &   39.029 & 0.2060 &   5.9 & 10.41  &      --  &   --   &  *  &      &   \\
{\bf A1413}&   178.826 &   23.408 & 0.1427 &  15.5 & 13.28  &    12.69 & 10.91  &  *  &    * & * \\
 A1423&   179.342 &   33.632 & 0.2130 &   5.3 & 10.03  &     3.73 &  7.23  &  *  &      & * \\
 A1682&   196.739 &   46.545 & 0.2260 &   5.3 & 11.26  &     4.07 &  8.79  &  *  &      &   \\
 A1758&   203.179 &   50.550 & 0.2799 &   3.6 & 11.68  &     5.60 & 18.29  &     &    * & * \\
 A1763&   203.818 &   40.996 & 0.2279 &   6.9 & 14.93  &     6.35 & 13.85  &  *  &    * &   \\
{\bf A1914}&   216.509 &   37.835 & 0.1712 &  15.0 & 18.39  &    12.90 & 15.91  &  *  &    * & * \\
 A2111&   234.924 &   34.417 & 0.2290 &   5.0 & 10.94  &     3.91 &  8.68  &  *  &      &   \\
{\bf A2218}&   248.970 &   66.214 & 0.1710 &   7.5 &  9.30  &     7.16 &  8.16  &  *  &    * & * \\
 A2219&   250.094 &   46.706 & 0.2281 &   9.5 & 20.40  &    11.18 & 24.33  &  *  &    * &   \\
 RXJ1720& 260.037 &   26.635 & 0.1640 &  14.3 & 16.12  &    10.84 & 12.34  &  *  &    * &   \\
 A2261&   260.615 &   32.127 & 0.2240 &   8.7 & 18.18  &     9.83 & 20.61  &  *  &    * &   \\
\end{tabular}
\caption{The selection of the cluster sample from the BCS and NORAS
surveys. Clusters are required to have $L_{\rm X} > 8 \times 10^{44}
\rm erg \, s^{-1}$ and $f_{\rm X} > 5 \times 10^{-12} \rm erg \,
s^{-1} \, cm^{-2}$ in both samples, and to be free of contaminating
radio sources. The final three columns indicate which clusters are
selected from BCS, NORAS, and radio data respectively. Eleven clusters
meet the X-ray selection criteria, and five of these (shown in bold)
are sufficiently free of radio sources to have their SZ effects
measured. Units of luminosity $L_{\rm X}$ are $ 10^{44} \rm erg \,
s^{-1}$ and of flux $f_{\rm X}$ are $10^{-12} \rm erg \, s^{-1} \,
cm^{-2}$.}
\label{table:sample}
\end{table*}

\begin{table*}
\begin{tabular}{lccccccc}
Cluster & $z$ & $\beta$ & $\theta_1$ ($^{\prime \prime}$) & $\theta_2$ ($^{\prime \prime}$)& PA $^{\circ}$& $n_0 \, h_{50}^{-1/2} \rm cm^{-3}$ & Temperature (keV)\\
\hline
A697&0.282&0.67&46 & 34 & 101 & $7.9 \times 10^{-3}$ & $8.0^{+0.55}_{-0.42}$\\
A773&0.217&0.64&61 & 44 & 16  & $6.8 \times 10^{-3}$ & $8.7^{+0.42}_{-0.42}$\\
A1413&0.143&0.58&56 & 38 & 1  & $9.6 \times 10^{-3}$ & $8.5^{+0.79}_{-0.48}$\\
A1914&0.1712&0.69&51&46 & 10 & $13.6 \times 10^{-3}$  & $8.6^{+0.36}_{-0.39}$\\
A2218&0.171&0.65&69&53 & $-8$ & $6.0 \times 10^{-3}$& $7.20^{+0.36}_{-0.36}$\\
\end{tabular}
\caption{X-ray parameters for the five clusters used in the
combined $H_0$ measurement. }
\label{table:xray}
\end{table*}

\begin{table}
\begin{tabular}{cccc}
Cluster & Flux density ($\mu$Jy) & RA offset (arcsec) & dec offset (arcsec)\\
\hline
A697 &  $ 462 $ & $-468 $ & $193$\\
     &  $ 154 $ & $-442 $ & $156$\\
     &  $ 215 $ & $-483 $ & $226$\\
     &  $ 624 $ & $28   $ & $-236$\\
\hline
A1914&    153 & $-262$ & $-195$\\
  &  379 & $-169$ & $-99$\\
  &  155 & $-141$ & $31$\\
  &  173 & $-108$ & $201$\\
  &  103 & $-118$ & $-52$\\
  &  278 & $-96$ & $-82$\\
  &  234 & $49$ & $66$\\
  &  123 & $-56$ & $210$\\
  &  186 & $13$ & $158$\\
\hline
A2218 &1438& $-189$ & 36\\
 &1227& $-40$ & 117\\
 &1541& $126$ & 94\\
 &522& $132$ & 105\\
 &543& $-196$ & 23\\
 &240& $-207$ & 41\\

\end{tabular}
\caption{Sources subtracted from the radio data. The flux densities
given are apparent, ie not corrected for the RT primary beam. Offsets
are from the radio pointing centres: A697 $\rm 08^h 42^m 57^s.7
+36^{\circ}.7 21^{\prime} 45^{\prime \prime}$; A1914 $\rm 14^h 26^m
02^s.0 +37^{\circ} 49^{\prime} 29^{\prime \prime}$; A2218 $\rm 14^h
26^m 02^s.0 +37^{\circ} 49^{\prime} 29^{\prime \prime} (J2000)$. The
errors on the source fluxes are $\pm 30$, $\pm 45$ and $\pm 35$
$\mu$Jy respectively.}
\label{table:sources}
\end{table}

\begin{table}
\begin{tabular}{lrrr}
Cluster & $H_0$ (SCDM) & $H_0$ ($\Lambda$CDM) & $\Delta T_0 \, (\mu \rm K)$\\
\hline
A697 & $52^{ +13}_{ -11}$  & $  59^{ +15}_{ -12}$ & 1047\\
A773 & $77^{ +19}_{ -17}$  & $  85^{ +20}_{ -19}$ & 737\\
A1413& $57^{ +23}_{ -16}$  & $  61^{ +25}_{ -17}$ & 863\\
A1914&$119^{ +46}_{ -38}$  & $ 129^{ +51}_{ -42}$ & 864\\
A2218& $31^{ +10}_{ -8 }$  & $  35^{ +11}_{ -9 }$ & 797\\
\hline
Sample& $59^{ +8}_{ -7 }$  & $  65^{ +8}_{ -7 }$ &\\
\end{tabular}
\caption{$H_0$ values for the five clusters in the sample, for both
SCDM and $\Lambda$CDM cosmologies. Also given are the central
temperature decrements using the best normalisation of the fitted
$\beta$ model to the SZ data.}
\label{table:h0}
\end{table}

\clearpage

\end{document}